\documentclass[aps,twocolumn,nofootinbib]{revtex4-1}
\usepackage{amssymb,amsmath,amstext,amsfonts,color,ulem,textcomp,amsthm}
\usepackage{graphicx,multirow,xcolor}

\newcommand{\beq}{\begin{equation}}
\newcommand{\eeq}{\end{equation}}
\newcommand{\bal}{\begin{aligned}}
\newcommand{\eal}{\end{aligned}}
\def\d{\mathrm {d}}

\begin{document}
	
\title{Geodesic completeness, cosmological bounces and inflation} 
	
\author{Sebastian Garcia-Saenz}
\email{sgarciasaenz@sustech.edu.cn}
	
\author{Junjie Hua}
\email{12112930@mail.sustech.edu.cn}
	
\author{Yunke Zhao}
\email{12210129@mail.sustech.edu.cn}
	
\affiliation{Department of Physics, Southern University of Science and Technology, Shenzhen 518055, China}

\begin{abstract}
		
The question of geodesic completeness of cosmological spacetimes has recently received renewed scrutiny. A particularly interesting result is the observation that the well-known Borde-Guth-Vilenkin (BGV) theorem may misdiagnose geodesically complete cosmologies. We propose a simple amendment to the BGV theorem which addresses such loopholes while retaining much of its generality. We give straightforward proofs of some recently offered conjectures concerning (generalized) Friedmann-Lema\^itre-Robertson-Walker spacetimes: geodesic completeness implies (i) the existence of a bounce, loitering phase or an emergent cosmology, and (ii) a phase of accelerated expansion with strictly increasing Hubble rate. Our results are purely kinematic and do not assume general relativity or energy conditions.
		
\end{abstract} 
	
\maketitle

\section{Introduction}

The problem of consistently embedding cosmic inflation within string theory has proved a very challenging task, as illustrated for instance by the `eta problem', by the difficulties to construct consistent de Sitter backgrounds in supergravity and by the so-called swampland conjectures (see \cite{Cicoli:2023opf,Palti:2019pca,Yamaguchi:2011kg,Baumann:2014nda} for reviews). One might expect a bottom-up approach to the issue to be a helpful supplement in this endeavor, analogously to the study of constraints on effective field theories based on well-tested, fundamental principles \cite{Adams:2006sv}. In the context of cosmology, and assuming a sufficiently low-energy regime such that a metric description of spacetime is applicable, one principle that one might naturally wish to impose is the absence of curvature singularities.

Inflation in this respect resolves the singularity problem of standard Big Bang cosmologies (provided a consistent description connecting inflation with the hot Big Bang phase), by extending the spacetime at early times with a phase of quasi-exponential expansion (using cosmic time for the time coordinate) which is manifestly free of singularities. Nevertheless, even this description is not fully satisfactory, since it is known that absence of singularities does not imply geodesic completeness, i.e.\ the property that geodesic curves may be extended without bound into the past. Inflation precisely fails this test, as may be seen more simply in the case of exact exponential expansion: the Friedmann-Lema\^itre-Robertson-Walker (FLRW) metric describes only a part of the full de Sitter spacetime, and (null and time-like) geodesics cannot be extended indefinitely into the past \cite{Hawking:1973uf,Vilenkin:1992uf}. A natural question is then whether inflationary cosmologies may be extended further into the past so as to achieve geodesic completeness \cite{Aguirre:2001ks,Geshnizjani:2023hyd,Yoshida:2018ndv,Nomura:2021lzz}.

A famous theorem by Borde, Guth and Vilenkin (BGV) appears to forbid any such extension, at least in the context of expanding spacetimes \cite{Borde:2001nh} (see also \cite{Borde:1993xh,Borde:1994ai} for earlier work and \cite{Kinney:2021imp,Kinney:2023urn,Kothawala:2018ghr,Conroy:2014dja,Vilenkin:2014yva} for related discussions and reformulations). More in detail, the BGV theorem states that if the average expansion rate (defined below) measured by an observer is positive, then the observer's geodesic must be past-incomplete.\footnote{We say that a (null or time-like) geodesic is past- or future-complete if it may be extended indefinitely into the past or future, respectively. A complete geodesic is both past- and future-complete. We say that a spacetime is past- or future-complete if all (null and time-like) geodesics are past- or future-complete, respectively.} The theorem is very general in that it does not assume the spacetime to be of the FLRW type, nor does it assume the field equations of general relativity or any energy conditions. However, an interesting recent study by Easson and Lesnefsky (EL) has shown that the original BGV theorem contains a subtle loophole and that, as a consequence, it may yield `false positives' when diagnosing past-incompleteness \cite{Lesnefsky:2022fen,Easson:2024uxe,Easson:2024fzn}.

We propose here a simple amendment to the BGV theorem which we claim closes all such loopholes. Our proposal maintains essentially all the simplicity (including its rather direct proof) and generality of the original, being still applicable to generic spacetimes and to any metric theory of gravity. As an application, we demonstrate that past-complete cosmologies must exhibit a phase of accelerated expansion with strictly increasing Hubble rate. As an additional result, in the particular context of generalized FLRW spacetimes, we establish that past-complete spacetimes must either experience a bounce, a loitering phase, or else be emergent (see below for the precise definitions).

\section{BGV theorem and its loophole}

We begin by recollecting the original formulation of the BGV theorem as well as some definitions that we shall make use of in what follows. Consider a spacetime $(\mathcal{M},g)$ and let $u^{\mu}$ be a time-like vector field such that the integral curves of this field define a future-oriented time-like geodesic congruence. Let $\mathcal{C}(\lambda)$ be a time-like or light-like geodesic curve (the `observer') with tangent $v^{\mu}$ and affine parameter $\lambda$. We define the expansion rate of the congruence, or `generalized Hubble parameter', with respect to the geodesic $\mathcal{C}$ as
\beq
H(\lambda)\equiv\frac{v_{\mu}v^{\nu}\nabla_{\nu}u^{\mu}}{v^2+\gamma^2} \,,
\eeq
where $v^2\equiv v^{\mu}v_{\mu}=-1$ or 0 respectively for the time-like and null cases (we use the mostly plus metric signature), $\gamma\equiv -v^{\mu}u_{\mu}$, and $u^{\mu}$ in this definition is taken to be evaluated on $\mathcal{C}(\lambda)$, i.e.\ $u^{\mu}=u^{\mu}(x(\lambda))$. One can check that this definition coincides with the usual Hubble parameter when applied to a geodesic congruence composed of comoving test particles in the context of generalized FLRW spacetimes. A \textit{generalized} FLRW spacetime is defined by a warped product metric of the form $\d s^2=-\d t^2+a^2(t)\d\Sigma^2$ in terms of cosmic time $t$ (the gauge we adopt throughout this paper), where $\d\Sigma^2$ is the line element of a singularity-free Euclidean manifold, not necessarily maximally symmetric.

Define next the average expansion rate by
\beq
H^{\mathcal{C}}_{\rm av}\equiv \frac{1}{\lambda_f-\lambda_i}\int^{\lambda_f}_{\lambda_i}\d\lambda \, H(\lambda) \,,
\eeq
where $\lambda_f$ is a reference value of the affine parameter, and $\lambda_i$ is taken as the infimum of the parameter for the geodesic under consideration. The geodesic is past-incomplete if $\lambda_i>-\infty$; otherwise it is past-complete.

It is straightforward to prove that \cite{Borde:2001nh}
\beq \label{eq:Hav identity}
H^{\mathcal{C}}_{\rm av}=\frac{F(\lambda_f)-F(\lambda_i)}{\lambda_f-\lambda_i} \,,
\eeq
where
\beq
F(\lambda)=\begin{cases}
	\dfrac{1}{\gamma} & \mbox{if $\mathcal{C}$ is light-like}\,, \\
 \\
	\dfrac{1}{2}\log\dfrac{\gamma+1}{\gamma-1} & \mbox{if $\mathcal{C}$ is time-like}\,. \\
\end{cases}
\eeq
Notice that $\gamma>0$ for null geodesics and $\gamma>1$ for time-like geodesics (since both $u^{\mu}$ and $v^{\mu}$ are future-oriented and not parallel), hence $F(\lambda)>0$ for all $\lambda\in(\lambda_i,\lambda_f)$. For a comoving geodesic congruence in a generalized FLRW spacetime with scale factor $a(t)$, one has $\gamma(\lambda)=1/a(t(\lambda))$ in the light-like case and $\gamma(\lambda)=\sqrt{a^2(t(\lambda))+c^2}/a(t(\lambda))$ in the time-like case ($c$ is a non-zero integration constant).

\textbf{Theorem 1 (original BGV).} If $H^{\mathcal{C}}_{\rm av}>0$, then the geodesic $\mathcal{C}$ is past-incomplete.

\textit{Proof.} Since $F(\lambda)>0$, one has from \eqref{eq:Hav identity} and the assumption that
\beq
0<H^{\mathcal{C}}_{\rm av}<\frac{F(\lambda_f)}{\lambda_f-\lambda_i} \,,
\eeq
and therefore $\lambda_i>-\infty$.$\qed$

EL have proposed a simple example that shows the flaw in the BGV theorem as given in the above formulation \cite{Easson:2024uxe}. A generalized FLRW metric with scale factor given by
\beq
a(t)=a_0e^{2t/\alpha}+c \,,
\eeq
with positive constants $a_0$, $\alpha$ and $c$, may be proved to be geodesically complete. The violation of the original BGV theorem may be seen by explicitly calculating $H^{\mathcal{C}}_{\rm av}$. For simplicity we consider here a light-like geodesic and choose $a_0=\frac{1}{2}=c$ and $\alpha=1$. We then have $\lambda(t)=\frac{1}{4}\left(e^{2t}+2t\right)$, modulo an arbitrary additive constant, and
\beq
H^{\mathcal{C}}_{\rm av}=\frac{2(e^{2t_f}-e^{2t_i})}{e^{2t_f}-e^{2t_i}+2(t_f-t_i)} \,,
\eeq
where $t_f\equiv t(\lambda_f)$ and $t_i\equiv t(\lambda_i)$. We see that $H^{\mathcal{C}}_{\rm av}$ is manifestly positive for all $t_i\in (-\infty, t_f)$, and yet the spacetime is geodesically complete, in contradiction with Theorem 1.

The trouble here is that the theorem assumes that the initial parameter $\lambda_i$ can be taken to be `strictly equal' to $-\infty$, which of course yields an ill-defined expression
for $H^{\mathcal{C}}_{\rm av}$. A complete geodesic is defined by the range $\lambda\in(-\infty,\infty)$, therefore if $H^{\mathcal{C}}_{\rm av}>0$ for all $\lambda_i\in(-\infty,\lambda_f)$, then it does not follow that the geodesic is past-incomplete. Other related issues of the original BGV theorem are discussed in Ref.\ \cite{Lesnefsky:2022fen}.

A different theorem to judge the geodesic completeness of a generalized FLRW spacetime has been given by EL \cite{Easson:2024fzn,Easson:2024uxe} (based on earlier works \cite{Sanchez1,Sanchez2}). We repeat here the theorem but focusing only on past-incompleteness of null and time-like geodesics.

\textbf{Theorem 2.} Let $(\mathcal{M},g)$ be a generalized FLRW spacetime with scale factor $a(t)$. The spacetime is past null complete iff the integral $\int_{-\infty}^{t_0}\d t\, a(t)$ diverges for all $t_0\in\mathbb{R}$. The spacetime is past time-like complete iff the integral $\int_{-\infty}^{t_0}\d t\, \frac{a(t)}{\sqrt{a^2(t)+1}}$ diverges for all $t_0\in\mathbb{R}$ \cite{Easson:2024fzn,Easson:2024uxe}.

We may check this theorem by studying some examples. Consider first the EL model given above. It is easy to check that indeed both integrals are divergent for any $t_0$, confirming that this model is geodesically complete. Consider second the simplest inflationary model with $a(t)=e^{t/\alpha}$. Then $\int_{-\infty}^{t_0}\d t\, a(t)=\alpha e^{t_0/\alpha}$ and $\int_{-\infty}^{t_0}\d t\, \frac{a(t)}{\sqrt{a^2(t)+1}}=\alpha\sinh^{-1}(e^{t_0/\alpha})$ are both convergent and the spacetime is past-incomplete. Indeed, this spacetime is of course nothing but the expanding patch of de Sitter spacetime alluded to above.

\section{Amendment to the BGV theorem}

We propose here an amended version of the BGV theorem.

\textbf{Theorem 3 (amended BGV).} Assume there exists $\Delta>0$ such that
\beq
H^{\mathcal{C}}_0\equiv \frac{1}{\lambda_f-\lambda_0}\int^{\lambda_f}_{\lambda_0}\d\lambda \, H(\lambda) \geq \Delta \,,
\eeq
for all $\lambda_0\in(\lambda_i,\lambda_f)$. Then the geodesic $\mathcal{C}$ is past-incomplete. Note that $\Delta$ may in general depend on the reference parameter $\lambda_f$.\footnote{After the submission of the first version of this work to the arXiv we became aware of Ref.\ \cite{Pavlovic:2023mke}, where the amended version of the BGV theorem was also given. We thank Petar Pavlovic for bringing his work to our attention.}

\textit{Proof.} As in the original proof, we have here the identity
\beq \label{eq:H0 identity}
H^{\mathcal{C}}_0=\frac{F(\lambda_f)-F(\lambda_0)}{\lambda_f-\lambda_0} \,.
\eeq
From the hypothesis and the fact that $F(\lambda)>0$ we then obtain
\beq \label{eq:H0 ineq}
\Delta\leq H^{\mathcal{C}}_0<\frac{F(\lambda_f)}{\lambda_f-\lambda_0} \quad\Rightarrow\quad \lambda_0>\lambda_f-\frac{F(\lambda_f)}{\Delta} \,.
\eeq
Since this holds for any $\lambda_0\in(\lambda_i,\lambda_f)$, it follows that $\lambda_i\geq \lambda_f-\frac{F(\lambda_f)}{\Delta}$.$\qed$

Consider again the EL model with the choice $a_0=\frac{1}{2}=c$ and $\alpha=1$ as a first example. Then we compute, in the light-like case,
\beq
H^{\mathcal{C}}_0=\frac{2(e^{2t_f}-e^{2t_0})}{e^{2t_f}-e^{2t_0}+2(t_f-t_0)} \,.
\eeq
Since $\lim_{t_0\to-\infty}H^{\mathcal{C}}_0=0$, we conclude that the required $\Delta$ does not exist, so the amended BGV theorem does not lead to the wrong conclusion that this model is past-incomplete. The same may be checked to hold in the time-like case.

In the case of expanding de Sitter, $a(t)=e^{t/\alpha}$, we have $H=1/\alpha$ and hence trivially $H^{\mathcal{C}}_0=1/\alpha\equiv\Delta$. As a less trivial example, consider the FLRW spacetime with scale factor $a(t)=-(\alpha/t)^3$, restricted to $t<0$ (this model is of course not future-complete; here we only wish to illustrate how to diagnose past-incompleteness using our theorem). This yields, in the light-like case,
\beq
H^{\mathcal{C}}_0=-\frac{2\left(t_0^2+t_0t_f+t_f^2\right)}{t_0t_f(t_0+t_f)} >-\frac{2}{t_f}\quad \forall t_0<t_f \,,
\eeq
indicating past-incompleteness, as may be also directly checked from Theorem 2.

As a verification of the amended BGV theorem, but specifically in the context of generalized FLRW spacetimes, we may show that the assumptions of the theorem imply the convergence of the integrals in Theorem 2, thus implying geodesic incompleteness. Rewriting Eq.\ \eqref{eq:H0 ineq} we have
\beq
0<\lambda_f-\lambda_0\leq \frac{F(\lambda_f)}{\Delta} \,.
\eeq
These inequalities directly translate into bounds, and therefore convergence, of the integrals in the EL theorem. Indeed these integrals are nothing but the range of the affine parameter $\lambda$, up to affine transformations \cite{Easson:2024fzn}.

More explicitly, for a light-like geodesic we have $\d\lambda =a(t)\d t$, up to a multiplicative constant (which must be positive since the geodesic is future-oriented). Therefore
\beq
0<\int_{t_0}^{t_f}\d t\, a(t)\leq \frac{F(\lambda_f)}{\Delta} \,.
\eeq
We now argue by contradiction. Assume the geodesic is past-complete, so that $\lambda_0\in(-\infty,\lambda_f)$, hence $t_0\in(-\infty,t_f)$. Therefore we may take the limit $t_0\to-\infty$ in the previous integral, and conclude its convergence, in contradiction with Theorem 2. Thus the geodesic must be past-incomplete. The proof for time-like geodesics is completely similar.

\section{Past completeness and bouncing cosmologies}

Our amended BGV theorem has an interesting application in the context of generalized FLRW spacetimes. As a preliminary result we show here that past-complete cosmological spacetimes must contain at least one bounce or else be emergent (see e.g.\ \cite{Ellis:2002we} and the reviews \cite{Novello:2008ra,Battefeld:2014uga,Brandenberger:2016vhg}). Here `bounce' is defined in the generalized sense introduced in Ref.\ \cite{Easson:2024fzn}, i.e.\ any time $t$ such that $\dot{a}(t)=0$, including the limiting cases where $\lim_{t\to\pm\infty}\dot{a}(t)=0$. This definition unifies several different scenarios: the usual notion of bounce, where $a(t)$ has a local minimum; a turning point corresponding to a local maximum of $a(t)$; a loitering point or phase where $\dot{a}(t)$ does not change sign but is momentarily zero; emergent cosmologies in which $a(t)$ approaches or is exactly equal to a constant in the asymptotic past. Actually, we will need here a more restricted definition of `emergent', namely that $\lim_{t\to-\infty}a(t)=c$, a non-negative constant. Note that this implies that $\lim_{t\to-\infty}\dot{a}(t)=0$ only if the latter limit exists. Similarly we say that a spacetime is `convergent' if $\lim_{t\to\infty}a(t)=c\geq 0$. The following lemma formalizes our result.

\textbf{Lemma.} Consider a generalized FLRW spacetime with smooth, non-constant scale factor $a(t)$. If the spacetime is geodesically complete, then it must exhibit at least one bounce or else be emergent or convergent.

\textit{Proof.} Suppose the spacetime is complete and does not have bounces, thus $\dot{a}(t)\neq0$ for all $t\in\mathbb{R}$. Therefore $a(t)$ is monotonic and bounded, hence it must converge either at $t\to\infty$ or $t\to-\infty$.$\qed$

This lemma is in agreement with the results of \cite{Carballo-Rubio:2024rlr} in the setting of general relativity. However it differs from (and in a sense trivializes) the conjecture made in \cite{Easson:2024fzn}, which claimed that complete generalized FLRW spacetimes must have a bounce. A counter-example to this is provided by
\beq
\dot{a}(t)=\sum_{n=1}^{\infty}e^{-n^4(t+n^2)^2} \,,
\eeq
and the scale factor is defined as the integral of this expression, choosing the integration constant such that $\lim_{t\to-\infty}a(t)>0$ (it is easy to verify that the limit exists). We are concerned here only with past-completeness so we restrict to $t<0$ (similar examples with full geodesic completeness may be easily constructed). It is straightforward to check, using Theorem 2, that this spacetime is past-complete, yet it does not have bounces: $\dot{a}(t)$ is strictly positive and $\lim_{t\to-\infty}\dot{a}(t)$ does not exist. This model is admittedly contrived, and it would be interesting to analyze the implications that such type of geometries may bear on the form of the matter content given a concrete gravitational theory, and if they may be judged inadmissible based on some physical criteria.

We also remark that in the emergent/convergent scenarios, the limit of $a(t)$ need not be strictly positive. As an example, consider $a(t)=-1/t+\tilde{a}(t)$, where $\tilde{a}(t)$ is the scale factor of the previous example, with $\lim_{t\to-\infty}\tilde{a}(t)=0$ (remember we restrict here to $t<0$). This model is past-complete, it does not have bounces, and it is emergent with $\lim_{t\to-\infty}a(t)=0$.

\section{Past completeness and inflation}

Our main result, also conjectured in \cite{Easson:2024fzn} in the context of generalized FLRW spacetimes, is that a past-complete cosmology must exhibit a phase of accelerated expansion. We can actually demonstrate a stronger result from which this property follows as a trivial corollary, namely that there must exist a phase with strictly increasing Hubble parameter. The following theorem formalizes this statement.

\textbf{Theorem 4.} Consider a generalized FLRW spacetime with smooth, non-constant scale factor $a(t)$. If the spacetime is past-complete, and if there exists $t_f$ such that $H(t_f)>0$, then there exists an interval $\mathcal{I}\subset(-\infty,t_f)$ such that $\dot{H}(t)>0$ for all $t\in\mathcal{I}$.

\textit{Proof.} If the spacetime has a bounce at some $t_0<t_f$, then the mean value theorem implies that there exists $t_m\in(t_0,t_f)$ such that $\dot{H}(t_m)>0$. By the assumption of smoothness, there must exist an open interval $\mathcal{I}$, with $t_m\in\mathcal{I}$, such $\dot{H}(t)>0$ for all $t\in\mathcal{I}$.

If there is no such bounce, then by the above Lemma the spacetime must instead be emergent, with $\dot{a}(t)>0$, hence $H(t)>0$, for all $t<t_f$. If there exists some $t_0<t_f$ such that $H(t_0)<H(t_f)$, then the conclusion follows using the same reasoning. Otherwise, i.e.\ if $H(t)\geq H(t_f)$ for all $t<t_f$, then we reach a contradiction to the assumption of past-completeness as per Theorem 3.$\qed$

We remark that the assumption that $H>0$ for some period of time is clearly needed to establish the theorem, since it is indeed easy to construct instances of past-complete spacetimes such that both $H(t)<0$ and $\dot{H}(t)\leq0$ for all $t$ (example: $a(t)=-t$, restricted to $t<0$).

Let us also observe that Theorem 4 admits a `generalized' formulation valid for any spacetime, not necessarily of the FLRW class. Indeed it suffices to assume that there exists a geodesic such that the spacetime is either `locally bouncing', in the sense that $H(\lambda_0)=0$ for some $\lambda_0<\lambda_f$, or else that it is `locally emergent', in the sense that $H(\lambda)>0$ for all $\lambda<\lambda_f$. It then follows that there exists an interval of the geodesic parameter in which $\d H(\lambda)/\d\lambda>0$.

\section{Discussion}

In this work we have primarily focused on the question of past-completeness as we think this aspect to be most physically relevant. However \textit{future}-completeness may be assessed in very similar terms. In fact, our Lemma requires full geodesic completeness. The condition of future-completeness is needed in order to rule out exceptions in which the scale factor is monotonically decreasing but, for instance, exhibits a singularity at some finite time (as in the example we discussed). An alternative, perhaps more physical version of the theorem, which only concerns with past-completeness, would necessitate the extra hypothesis that $\dot{a}>0$ for some time interval. Theorems 3 and 4 also admit `time-reversed' versions. Alternatively, we could reformulate Theorem 4 with the assumption of full geodesic completeness, but without assuming $H>0$ for some time interval. The result that $\dot{H}>0$ for some period would still follow. 

Theorem 4 is of clear importance to inflation, as it demonstrates that slow-roll models with $\epsilon\equiv -\dot{H}/H^2>0$ are inconsistent with geodesic completeness. An earlier phase with $\dot{H}>0$ is necessary, at least if one aims at a complete description of the pre-hot Big Bang epoch based on FLRW spacetime. In general relativity, $\dot{H}>0$ typically translates into a violation of the null energy condition (NEC). Thus our theorem leads to the conclusion that any model of the early universe that respects the NEC must, within general relativity, be either singular or geodesically incomplete. This is of course in agreement with standard singularity theorems \cite{Penrose:1964wq,Hawking:1970zqf}, although again our result is a priori more generally applicable.

Violations of the NEC are often associated with instabilities \cite{Libanov:2016kfc,Dubovsky:2005xd,Hsu:2004vr,Buniy:2005vh,Buniy:2006xf,Rubakov:2014jja} or, more generally, with energy densities unbounded from below in phase space \cite{Sawicki:2012pz}. On the other hand, at least as far as linear stability is concerned, there exist examples of stable models featuring violations of the NEC \cite{Creminelli:2016zwa,Creminelli:2006xe,Nicolis:2009qm,Creminelli:2012my,Nicolis:2011pv,Creminelli:2010ba,Easson:2011zy}. Intriguingly, the proposal of \cite{Creminelli:2010ba} is precisely characterized by an emergent, past-complete FLRW metric where the scale factor approaches a constant in the past and evolves with increasing Hubble parameter. It would be interesting to study this question outside the context of FLRW cosmology, for instance to consider anisotropic models. Since our proposal, Theorem 3, is applicable to any spacetime, we expect it to be useful in addressing this problem. One possible such avenue would be to explore the consequences of the `generalized' version of Theorem 4 for other classes of spacetimes. We leave this investigation for future work.

\vskip 5pt

\textit{Acknowledgments.---}We would like to thank Jun Zhang for an insightful discussion. This work received support from the NSFC Research Fund for International Scientists (Grant No.\ 12250410250).

\newpage
	
\bibliographystyle{apsrev4-1}
\bibliography{AmendedBGVbib}
	
\end{document}